\documentclass[twocolumn,showpacs,preprintnumbers,amsmath,amssymb]{revtex4}
\usepackage{graphicx}
\begin{document}

\title{Calogero-Sutherland-Lieb-Liniger gas in one-dimensional cold atoms}

\author{Yue Yu}
\affiliation{Institute of Theoretical Physics, Chinese Academy of
Sciences, P.O. Box 2735, Beijing 100080, China}
\date{\today}
\begin{abstract}
We study an array of cigar-like Bose atom condensates confined in
a cylinder and examine the competition between the dipole-dipole
and the short range interactions. The system is effectively
reduced to a one-dimensional boson one with a contact and inverse
square interactions. We call this system the
Calogero-Sutherland-Lieb-Liniger
 gas. The universal properties of the ground
state are analyzed by the renormalization group theory. By using
the bosonization techniques to the excluson gas, we calculate the
non-universal exponent depending on the microscopic parameters.
This exponent may be experimentally measurable.

\end{abstract}


 \maketitle

One-dimensional many-particle systems have been a central topics
in physics for four decades. Many exactly soluble models were
proposed after Bethe solved the one-dimensional Heisenberg model
by his famous ansatz \cite{bethe,yangy}. Among them two kinds of
the exactly models are cross-sectional: The particles interacting
with contact \cite{ll} or with inverse square potentials
\cite{ca,suth}. The one-dimensional cold alkali metal Bose atom
gas, which may condense at zero temperature \cite{ketterle}, is a
typical example of the former. Here, we call it the Lieb-Liniger
(L-L) gas for Lieb and Liniger first gave the exact solution for
this model. The strong interacting limit of the L-L model, known
as the Tonks-Girardeau (T-G) gas \cite{TG}, has also reached in
the ultracold Bose atoms \cite{TGE}. The model with inverse square
interaction, called the Calogero-Sutherland(C-S) models play a
fundamental role in the contemporary theoretical physics
\cite{csmb}. The C-S model may be a universal Hamiltonian of some
weakly disordered metals \cite{sla}. The models may also be used
to describe the edge excitations in the fractional quantum Hall
effect \cite{yu}. In a recent work \cite{yu1}, we proposed a
possible realization of the C-S gas for the cold atom with a
dipole-dipole interaction, which is based on a major progress in
the cold chromium atoms $^{52}$Cr i.e., this cold atom gas is
turned into a Bose condensate \cite{dpdp}. It may also be
realizable in the cold atom in tight waveguides with anyonic
exchange symmetry \cite{gi}.

The models with both the short range and long range interactions
mentioned above have been applied to the faceting transition of
the miscut crystal surface\cite{LN,lu} by renormalization group
analysis \cite{RG}. The low energy effective theory of these
models may be characterized by the non-ideal excluson gas
\cite{wuy} since the quasi-particle excitations in these models
obey exclusion statistics \cite{halds,wu,wub}. It is a Luttinger
liquid theory \cite{hald,lutt} whose non-universal exponent is
given by both the short range interaction strength and the
exclusion statistics parameter.

\begin{figure}%[htb]
\begin{center}
\includegraphics[width=6.0cm]{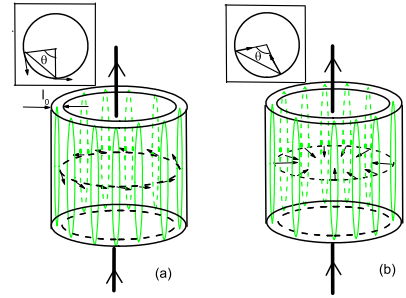}
\end{center}
%\vspace{-1cm}
 \caption{\label{fig1} (Color on-line)
 The cigar-like condensates are confined in a cylinder. (a) The magnetic
 dipoles; (b) The electric dipoles. The insets show the angle relation in
 deriving eq. (\ref{3})}.
%\vspace{-0.7cm}
\end{figure}

Both interactions are of the comparative magnitude in the cold
chromium atom gas \cite{dpdp}. This offers an opportunity to study
the intergradation from one to another since both interactions are
adjustable \cite{fesh,adipole}. In this Letter, we propose a
realization of such a model for the cold atoms in a cylinder. Fig.
\ref{fig1} is a sketch map of our set up. An array of cigar-like
Bose atom condensates is confined in a cylinder which is made by,
e.g., a ring-shaped optical lattice \cite{rsl}. The atoms interact
with a contact potential accompanied a dipole-dipole potential
\cite{yy}. An electromagnetic field is generated by a stable
current. If the dipoles is magnetic, their orientations are shown
in Fig. \ref{fig1}(a) while the orientations are as in Fig.
\ref{fig1}(b) if they are electric. The arc coordinate is $x$ and
the vertical direction is $y$. The two-dimensional wave function
of the many atom system is factorized, i.e., the $y$-direction
wave function is approximated by its condensate wave function and
the system  effectively is one-dimensional. The effective atom
density of the system may be written as $\rho({\bf
r})=\rho_0(y)\sum_{i=1}^N\delta(x-x_i)$ where $\rho_0(y),$
approximated by a constant $\rho_0$, is the atom density in a
single cigar-like condensate and $N$ the number of the
condensates.

The three-dimensional interacting potential of the atom cloud is
generally given by $ V=\frac{1}2\int_S d{\bf r}d{\bf r'}\rho({\bf
r})V({\bf r}-{\bf r}')\rho({\bf r}')$ where $S$ denotes the
cylinder with the thick $l_0$. For simplicity, we consider a
dilute gas limit in which whatever an optical lattice exists, the
continuous model works. The dilute condition is given by
$r_0N/L<<1$ where $L$ is the perimeter of the circle and $r_0$ is
the azimuthal width of a condensate. The interaction $V({\bf
r}_{ij}={\bf r}_i-{\bf r}_j')$ consists of two parts
\begin{eqnarray}
V=\sum_{i<j}\biggl[U\delta({\bf r}_{ij})+\frac{{\bf d}_i\cdot {\bf
d}_j-3({\bf d}_i\cdot \hat{\bf r}_{ij})({\bf d}_j\cdot \hat{\bf
r}_{ij})}{r_{ij}^3}\biggl],
\end{eqnarray}
where $U=\frac{4\pi \hbar^2 a}m$ for $a$ is the $s$-wave
scattering length. ${\bf d}$ is the dipolar moment of the atom.
For the factorized wave function, the Hamiltonian may be reduced
to a one-dimensional one
\begin{eqnarray}
H_1&=&-\sum_{i=1}^N\frac{\hbar^2}{2m}\frac{d^2}{d
x_i^2}+U_1\sum_{i<j} \delta_{r_c}(x_{ij})\label{3}\\
&+&\frac{G\pi^2}{L^2}\sum_{i<j}\sin^{-2}\frac{\pi(x_i-x_j)}L
-\delta_\pm\frac{g\pi^2}{L^2}N(N-1)\nonumber
\end{eqnarray}
 where $G=
2d^2\rho_0$, $U_1\propto\frac{\hbar^2a}{ml_0 r_0}$ is the
one-dimensional reduction of $U$, whose detailed version depends
on the constraint potential. $r_c\to 0$
 is the character size of the contact potential.
$\delta_\pm=\frac{1\pm 2}2$ and the sign $\mp$ corresponds to the
magnetic or electric dipoles in Fig.\ref{fig1}(a) or (b). The
$\delta_\pm$ term in the Hamiltonian is not important in the
thermodynamic limit because it is of order $O(1)$ while the ground
state energy of the system is an extended quantity. We call this
system the L-L-C-S gas since it combines the C-S  and L-L gases.
In the following, we use the dimensionless coupling constants $
g=\frac{m}{\hbar^2} G,~~u=\frac{m r_c}{\pi\hbar^2}U_1,~{\rm and}~
v=u-g$ with $r_c /l_0$ fixed when $r_c$ and  $l_0\to 0$ .In the
system our are concerning, $g\geq 0$ while $u$ may be a real
number.

\begin{figure}%[htb]
%\vspace{-0.2cm}
\begin{center}
\includegraphics[width=6.0cm]{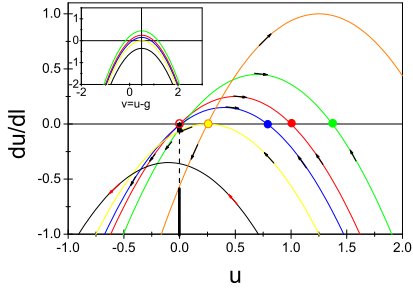}
\end{center}
%\vspace{-1cm}
 \caption{\label{fig2} (Color on-line) The renormalization group flows
 for $g$=0.75 (orange), 0.2 (green), 0 (red), -0.1 (blue),  -0.25 (yellow) and
 -0.6 (black). The arrows show the flow directions. The filled
 spots are the stable fixed points and the empty are the unstable
 fixed points.
 }
 %\vspace{-0.3cm}
\end{figure}

\begin{figure}%[htb]
%\vspace{-0.2cm}
\begin{center}
\includegraphics[width=6.5cm]{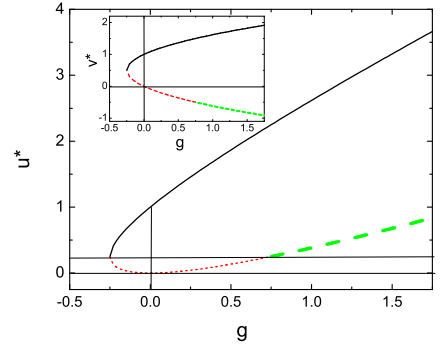}
\end{center}
%\vspace{-1cm}
 \caption{\label{fig3} (Color on-line) The fixed point lines. The
 solid(black) curve is the stable fixed points. In the red curves
 a pair of different $g$ are corresponding to the same unstable
 fixed point $u^*$. And in the green, the unstable fixed point
 for $g>0.75$ shares the same $u^*$ with the stable fixed point.
 }
 %\vspace{-0.3cm}
\end{figure}

To understand the universal properties of the ground state, we
employ the renormalization group theory. This has been built for
models with both contact and long range interactions \cite{RG}.
For this inverse square model in the one-dimension, the
renormalization group equation reads
\begin{eqnarray}
\frac{d u}{dl}=(2g+1)u-u^2-g^2,~{\rm or}~\frac{d v}{dl}=v-v^2+g
\end{eqnarray}
while the long range coupling constant $g$ is a marginal operator.
Here $l=\ln N$ is the running parameter. The flow pattern $\frac{d
v}{dl}$ and the fixed point values $v^*$ are shown in the insets
of Fig. \ref{fig2} and Fig. \ref{fig3}, respectively \cite{LN,RG}.
Since $g\geq 0$ in the present model, one may write
$g=\lambda(\lambda-1)$ for a real $\lambda$. The fixed point value
$v^*$ for a given $g$ equals to $\lambda$. In a pure C-S model, it
is known that $\lambda$ serves as the exclusion statistics
parameter\cite{wub}. This means that the fixed point line of this
model is determined by exclusion statistics of the system. To
understand the physics directly from the microscopic parameters,
we redraw the flow and fixed point line in $u-\frac{du}{dl}$ and
$g-u^*$ in Fig. \ref{fig2} and Fig. \ref{fig3}, respectively. In
Fig. \ref{fig2}, we see that there are two fixed points in a flow
line for a given $g>-0.25$. The right hand side one is stable and
the left one is unstable. Each fixed point value $u^*$ corresponds
to two $g_\pm=u^*\pm\sqrt{u^*}$. If $0<u^*<0.25$, both fixed
points are unstable. For $u^*>0.25$, one of $g_+$ is unstable
while $g_-$'s is stable. Fig. \ref{fig3} clearly shows this fact.
It is seen that there is no fixed point with $u^*<0$ or $g<-0.25$.
If $g<-0.25$, the attraction between condensates (called particles
hereafter) is so strong that the on-site repulsion can not
withstand the collapse. This collapse, however, will not happen
for the system we are considering, in which $g>0$.

If we only concentrate on the short range interaction, the above
renormalization group analysis for the ground state properties may
be complete. However, since the long range interaction may lead to
consequences with the exception of the above renormalization group
analysis, we have to carefully examine the ground state. The
renormalization group flows show that the pure C-S model with
$u=0$ is driven to $u\to -\infty$. We examine the meaning of this
running now. There are two real solutions for
$g=\lambda(\lambda-1)$ if $g>-0.25$, i.e.,
$\lambda_\pm=0.5(1\pm\sqrt{1+4g})$. The ground state energy is
given by \cite{suth} $
E_\lambda=\frac{\pi^2\lambda^2}{L^2}\sum_{j=1}^N(2j-N-1)^2$.
 For
$-0.25<g<0$, $0<\lambda_-<0.5$ and the ground state wave function
of the C-S model is given by
\begin{eqnarray}
\Psi_{C-S}=\prod_{i<j}\biggl[\sin\frac{\pi(x_i-x_j)}L\biggr]^{\lambda_-}.
\label{ln}
\end{eqnarray}
This means that although the interaction is attractive, there is
still a mutual repulsion between particles caused by the quantum
zero motion such that the system is still stable against the
collapse. Since the contact potential is less singular than the
$1/x^2$ potential, for $-0.25<g<0$, no matter what is the
magnitude of $u$, the ground state wave function exactly given by
(\ref{ln}). Reflected in Fig. \ref{fig2} , there is a small region
between the red curve and the yellow curve at $u=0$ (the black
short stub), the system may be characterized by the C-S model.  We
have discussed such a situation in our previous work \cite{yu1}
and do not concern it here because in our model $g>0$.

For $g>0.75$, $\lambda_-<-0.5$, and (\ref{ln}) is not square
integrable, the ground state wave function is given by \cite{suth}
\begin{eqnarray}
\Psi_{C-S}=\prod_{i<j}\biggl[\sin\frac{\pi(x_i-x_j)}L\biggr]
^{\lambda_+},\label{lp}
\end{eqnarray}
with $\lambda_+>1.5$. Hence, the system will not run to
$u=-\infty$ from the unstable fixed points but is characterized by
the thick vertical line at $u=0$ in Fig. \ref{fig2}.

For $0<g<0.75$ ($-0.5<\lambda_-<0$), there is a ground state
(\ref{ln}) which is square integrable. The renormalization group
flow indeed will drive the system to a state with $u=-\infty$ for
a negative $u$ fluctuation. However, at exact $u=0$, because of
the symmetry $\lambda\leftrightarrow-\lambda$ \cite{kaw}\cite{lu},
the system is equivalent to that with $0<\lambda<0.5$. This
character of the system is shown by the vertical dash line in Fig.
\ref{fig2}.

According to the above analysis, we have an overall view to the
ground states of the system. We see that there is no  ground state
with $0.5<\lambda<1.5$ in the present model. This is because the
non-interaction particles are boson, the system has to back to the
free boson when $g=u=0$. If applying this model to fermions,
$\lambda_-$ solutions have to be abandoned because $\lambda\to 1$
as $g\to 0$.

In the cold atom context, many system parameters may be exactly
controlled. Therefore, one should do microscopic calculations to
understand many non-universal properties. Here we use an effective
theory early developed in \cite{wuy}, to study some low energy
behaviors of the system. The theory was called the excluson gas
theory. Due to the symmetry $\lambda\leftrightarrow-\lambda$, we
focus on $\lambda\geq0$.

\begin{figure}%[htb]
%\vspace{-1.5cm}
\begin{center}
\includegraphics[width=8.0cm]{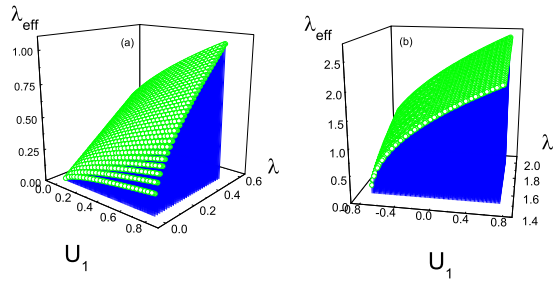}
\end{center}
%\vspace{-1.5cm}
 \caption{\label{fig4} (Color on-line) The effective Luttinger
 liquid parameter $\lambda_{eff}$(the green surfaces) varies as $\lambda$ and $U_1$.
 The unit of $U_1$ is $\hbar v_s$. (a)
 $0<\lambda<0.5$; (b) $\lambda>1.5$. }
 %\vspace{-0.3cm}
\end{figure}

We first consider $\lambda\ne 0$.  The pseudo-Fermi-momentum is
defined by $k_F\equiv \sqrt{2m\mu}$ for the chemical potential
$\mu$ . Its value is fixed by $\int_{-k_F}^{k_F} dk
\rho(k)=\frac{N_0}L\equiv\bar d_0$ for $N_0$ is the
one-dimensional particle
 number in the ground state. The
particle density $\rho(k)=1/(2\pi\lambda\hbar)$ if $|k|<k_F$ and 0
if $|k|>k_F$. This gives $k_F=\pi\lambda\hbar\bar d_0$ and
$\mu=\frac{1}{2m}(\pi\lambda\hbar\bar d_0)^2$. The ground state
energy can be written as, in the thermodynamic limit,
$E_\lambda/L=\frac{1}{2m}\int_{-k_F}^{k_F}dk \rho(k)
k^2=\frac{\hbar^2}{6m}\pi^2\lambda^2 \bar d_0^3$. the low-lying
excitations of the system are: The density fluctuations with the
sound velocity $v_s=v_F=k_F/m$; Adding extra number $M=N-N_0$ of
particles to the ground state; And creating a persist current by
the Galileo boost $k\to k+\pi J\hbar/L$ in the ground state. To
obey the periodic condition, on requires $(-1)^J=(-1)^M$ for
integers $M$ and $J$. If we consider only these low-lying
excitations, the non-ideal excluson gas may be described by a
bosonized effective Hamiltonian \cite{wuy}
\begin{eqnarray}
H_B&=&\frac{1}{2}\sum_{q>0}q\biggr[(v_s+U_1/\hbar)(b_q^\dagger b_q
+\tilde{b}_q^\dagger \tilde{b}_q+b_qb_q^\dagger
+\tilde{b}_q\tilde{b}_q^\dagger) \nonumber\\
&+&(U_1/\hbar)(b_q^\dagger \tilde{b}^\dagger_q +b_q
\tilde{b}_q+\tilde{b}^\dagger_qb_q^\dagger +\tilde{b}_qb_q)\biggr]
\label{bosonH}\\
&+&\frac{\pi}{L}[U_1(M_R^2+M_L^2)+2U_1M_RM_L]\nonumber\\
&+&\frac{\hbar}{2}\frac{\pi}{L} [v_N M^2+v_J J^2]\nonumber,
\end{eqnarray}
where $v_N=\lambda v_s$ and $v_J=v_s/\lambda $ are the standard
Luttinger liquid relation \cite{hald}. The boson operators $b_q$
and $\tilde b_q$ are related to the density fluctuating operators
$\rho_q^{(\pm)}$ by $b_q=\sqrt{2\pi/
qL}\,\rho_q^{(+)},~~\tilde{b}_q=\sqrt{2\pi/ qL}\,
{\rho}_q^{(-)\dagger}$ with $\pm$ denoting the right and left
Fermi points. According to the periodic boundary condition, a
consistent choice for $M=M_R+M_L$ and $J=J_R+J_L$ is
$M_R=J_R,M_L=-J_L$. Using the Bogoliubov transformation, the
Hamiltonian can be easily diagonalized
\begin{eqnarray}
H_B=\sum_{q>0}\tilde v_s q(a_q^\dagger a_q
+\tilde{a}_q^\dagger\tilde{a}_q)
+\frac{\hbar\pi}{2L}[\tilde{v}_NM^2
+\tilde{v}_JJ^2],\label{bosonHd}
\end{eqnarray}
where the Bogoliubov boson is given by
$a_q^\dagger=b^\dagger_q\cosh\tilde{\varphi}_0 -
\tilde{b}_q^\dagger\sinh\tilde{\varphi}_0,~
\tilde{a}_q^\dagger=\tilde{b}^\dagger_q\cosh\tilde{\varphi}_0
-b_q^\dagger\sinh\tilde{\varphi}_0$. The renormalized velocities
are $\tilde{v}_s= |(v_s+U_1)^2-U_1^2|^{1/2},~ \tilde{v}_N=
\tilde{v}_s e^{-2\tilde{\varphi}_0},~{\rm and
}~\tilde{v}_J=\tilde{v}_s e^{2\tilde{\varphi}_0}
$
with the controlling parameter $\tilde{\varphi}_0$
 determined by
\begin{equation}
\tanh(2\tilde{\varphi}_0)
=\frac{v_J-v_N-2U_1/\hbar}{v_J+v_N+2U_1/\hbar}.
\end{equation}
Thus, the Luttinger liquid relation survives with the effective
Luttinger liquid parameter
\begin{equation}
\lambda_{eff} = e^{-2\tilde{\varphi_0}}. \label{eff}
\end{equation}
There is a restriction in $U_1$  for a given $\lambda$ because
$|\tanh 2\tilde\phi_0|<1$. For $\lambda>0$, this holds if
$U_1>-\frac{1}2\lambda\hbar v_s$. Fig. \ref{fig4} shows the
Luttinger liquid parameter changes as interaction. An attractive
$U_1$ suppresses $\lambda_{eff}$ from $\lambda$ while a repulsive
one lifts $\lambda_{eff}$ as expected. This exponent
$\lambda_{eff}$ may determine the non-universal exponents in the
Luttinger liquid \cite{hald}. For example, the asymptotic single
particle correlation function is proportional to
$x^{-\lambda_{eff}}$ so that the momentum distribution is given by
$n(p)\sim C_1-C_2{\rm sgn}(p-p_F)|p-p_F|^{(\lambda_{eff}-1)}$. In
the limit $U_1\to \infty$, the system goes an insulator state
because $x^{-\lambda_{eff}}\to 0$ as $\lambda_{eff}\to \infty$.

\begin{figure}%[htb]
%\vspace{-1.5cm}
\begin{center}
\includegraphics[width=5.5cm]{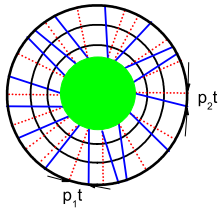}
\end{center}
%\vspace{-1.5cm}
 \caption{\label{fig5} (Color on-line) The sketch map of the image of the
 time-of-flight.  Circles represent the positions of the
 atoms with the fastest radial velocity at the different flight time.
 The cross points between each blue solid lines and the circle
 is the position of these atoms. The red dotted lines are the reference
 radial lines.}
 %\vspace{-0.3cm}
\end{figure}

The non-ideal gas theory does not work for the pure Lieb-linger
gas because eq.(\ref{eff}) is back to the free boson when
$\lambda\to 0$. For $\lambda=0$, the model is exactly soluble. We
can use the excluson gas theory with mutual statistics which was
also developed in ref. \cite{wuy}. It is a bosonization theory
with a similar Hamiltonian to (\ref{bosonHd}) with the effective
Luttinger liquid parameter
$\lambda_{eff}=[z(k^{(+)}_F)z(k^{(-)}_F)]^{-1}$ where $z(k)$ is
given by the integration equation $
z(k)=1+\int_{-k_F}^{k_F}\frac{dk'}{2\pi}\frac{U_1/2}{(U_1/2)^2+(k-k')^2}
z(k'). $ The pseudo-Fermi momentum $k_F$ is determined by the
vanishing point of the dressed energy \cite{yangy,wuy} and
$k_F^{(\pm)}=k_F\pm\epsilon$ for $\epsilon\to 0^+$. In the T-G
limit ($U_1\to \infty$), one finds $z(k^{(+)}_F)=1$ and
$z(k^{(-)}_F)=2$, i.e., $\lambda_{eff}=1/2$ \cite{ll}.

Now, we discuss the possible experimental implication.
$\lambda_{eff}-1$ is the exponent of the one-dimensional momentum
distribution, which can be measured by the absorption image of the
time-flight. Keeping the inner wall potential of the cylinder and
switching off the outer wall potential, the atom cloud will freely
expand. The atoms with the fastest radial velocity will run along
the blue lines in Fig. \ref{fig5} if they have the different
azimuthal momenta $p$. Counting the number of the different arc
intervals ($pt$) at a given time $t$, one can draw out the
momentum distribution of the cigar-like condensates before
releasing. Then the exponent $\lambda_{eff}$ can be determined.
The system with $\lambda\ne0$ is very different from $\lambda=0$.
This can be seen from the limit $U_1\to \infty$. For the former,
$\lambda_{eff}\to \infty$ and the system is an insulator while the
latter has $\lambda_{eff}=1/2$, the T-G gas state.

I thank S. M. Bhattacharjee to draw my attention to the
renormalization group analysis to the C-S-L-L model.
 This work was supported in part by Chinese
National Natural Science Foundation.

%\vspace{1cm}


\begin{references}
\bibitem{bethe} H. Bethe, Z. Physik {\bf 71}, 205 (1931).
\bibitem{yangy} C. N. Yang and C. P. Yang, J. Math. Phys. {\bf 10}, 1115
(1969).

\bibitem{ll} E. H. Lieb and W. Liniger, Phys. Rev. {\bf
130}, 1605 (1963).
\bibitem{ca} F. Calogero, J. Math. Phys. {\bf 10}, 2191 and 2197 (1969).
\bibitem{suth} B. Sutherland, J. Math. Phys. {\bf 12}, 246 (1971); Phys. Rev.
A {\bf 4}, 2019 (1971); {\bf 5}, 1372 (1972).
\bibitem{ketterle} W. Ketterle and N. J. vanDruten, Phys. Rev. A
{\bf 54}, 656 (1996).
\bibitem{TG} L. Tonks, Phys. Rev. {\bf 50}, 955(1936); M.
Girardeau, J. Math. Phys. {\bf 1}, 516 (1960).
\bibitem{TGE} T. Kinoshita et al, Sciences {\bf
305}, 1125 (2004); B. Paredes et al, Nature {\bf 429},277 (2004).


\bibitem{csmb} See e.g., J. F. van Diejen and L. Vinet (Eds.),
{\it Calogero-Moser-Sutherland Models: CRM Series in Mathematical
Physics 2000 XXV}, (Springer, Berlin, Heidelberg, New York 2000)
and references therein.

\bibitem{sla}B. D. Simons, P. A. Lee, and B. L. Altshuler, Phys. Rev. Lett.
70, 4122 (1993).

\bibitem{yu} Y. Yu and Z. Y.  Zhu, Comm. Theor. Phys. {\bf 29},
 351 (1998); Y. Yu, W. J. Zheng and
Z. Y. Zhu, Phys. Rev. B {\bf 56}, 13279 (1997); W. J.  Zheng and
Y. Yu,  Phys. Rev. Lett. {\bf 79}, 3242 (1997).

\bibitem{yu1} Y. Yu, e-preprint, cond-mat/0603340.

\bibitem{dpdp} A. Griesmaier, J. Werner, S. Hensler, J. Stuhler,
 and T. Pfau, Phys. Rev. Lett. {\bf 94}, 160401(2005).

\bibitem{gi} M. D. Girardeau, e-print-cond-mat/0604357.

\bibitem{LN} R. Lipowsky and Th. M. Nieuwenhuizen, J. Phys. A {\bf 21},
L89 (1988). R. K. P. Zia, R. Lipowsky and D. M. Kroll, Am. J. Phys
{\bf 56}, 160 (1988). S. M. Bhattacharjee, Phys. Rev. Lett. {\bf
76}, 4568 (1996); {\bf 83}, 2374 (1999).

\bibitem{lu} M. Lassig, Phys. Rev. Lett. {\bf 77}, 526 (1996).

\bibitem{RG} E. B. Kolomeisky and J. P. Straley, Phy. Rev. B {\bf
46}, 13942 (1994), and references therein.

\bibitem{wuy} Y. S. Wu and Y. Yu, Phys. Rev. Lett. {\bf 75}, 890
(1995); Y. S. Wu, Y. Yu and H. X. Yang, Nucl. Phys. B {\bf 604},
551 (2001). Y. Yu, H. X. Yang and Y.S. Wu, e-preprint,
cond-mat/9911141.

\bibitem{halds} F. D. M. Haldane, Phys. Rev.
Lett. {\bf 67}, 937 (1991).
\bibitem{wu} Y. S. Wu, Phys. Rev. Lett.
{\bf73}, 922 (1994).
\bibitem{wub} D. Bernard and Y. S. Wu,
in Proc. 6th Nankai Workshop, eds. M. L. Ge and Y. S. Wu, World
Scientific (1995).




\bibitem{hald} F. D. M. Haldane, J. Phys. C {\bf 14}, 2585
(1981).
\bibitem{lutt} A. Luther and I. Peschel, Phys. Rev. B {\bf
9}, 2911 (1974); E. H. Lieb and D. C. Mattis, J. Math. Phys. {\bf
6}, 304(1965); J. M. Luttinger, J. Math. Phys. {\bf 4}, (1963); S.
Tomonaga, Prog. Theor. Phys. {\bf 5}, 544 (1950).

\bibitem{fesh} S. Inouye et al, Nature (London) {\bf 392}, 151 (1998);
J.L. Roberts et al, Phys. Rev. Lett. {\bf 86}, 4211 (2001).

\bibitem{adipole} S. Giovanazzi, A. G¡§orlitz, and T. Pfau, Phys. Rev. Lett.
{\bf 89}, 130401 (2002).


\bibitem{rsl} L. Amico et al, Phys. Rev. Lett.
{\bf 95}, 063201 (2005).

\bibitem{yy} S. Yi and L. You, Phys.
Rev. A {\bf 61}, 041604(R) (2000); L. Santos et al, Phys. Rev.
Lett. {\bf 85}, 1791 (2000); K. Goral et al., Phys. Rev. A {\bf
61}, 051601(R) (2000).


\bibitem{kaw} N. Kawakami and S. K. Yang, Phys. Rev. Lett. {\bf
67}, 2493 (1991).

\end{references}
\end{document}